\documentclass[aps,pre,twocolumn,floatfix]{revtex4-2}
\usepackage{bm}
\usepackage{mathrsfs,amsbsy,amssymb,latexsym,amsfonts,amsmath,fancyhdr}
\usepackage{graphicx}
\usepackage{color}
\usepackage{url}
\usepackage[normalem]{ulem}
\usepackage[hidelinks]{hyperref}
\usepackage{mhchem}
\usepackage{cases}

\newcommand{\bv}[1]{{\boldsymbol #1}}

\newcommand{\kB}{k_{\rm B}}

\newcommand{\ep}{\epsilon}

\newcommand{\subG}{\mathrm{G}}
\newcommand{\subL}{\mathrm{L}}
\newcommand{\subLG}{\mathrm{L/G}}
\newcommand{\subC}{\mathrm{c}}

\newcommand{\ps}{p_\mathrm{s}}

\newcommand{\inty}{\int_{-L_y/2}^{L_y/2} \!\! dy~}

\begin{document}
\title{Pressure-drop localization and momentum insulation in liquid-gas coexistence Poiseuille flow}

\author{Naoko Nakagawa}
\affiliation {Department of Physics, Ibaraki University, Mito 310-8512, Japan}

\author{Shin-ichi Sasa}
\affiliation {Department of Physics, Kyoto University, Kyoto 606-8502, Japan}

\date{\today}
\begin{abstract}
We study Poiseuille flow of a one-component fluid between adiabatic plates in liquid-gas coexistence with a stationary interface.
Using only bulk transport laws and particle and energy conservation, we identify small dimensionless
parameters $A^\subL$ and $A^\subG$ governed by squared microscopic-to-macroscopic length ratios.
In weak driving with macroscopic bulk regions and a regular interfacial temperature response, the pressure drop is concentrated across the interface, strongly suppressing the particle current.
For equal-temperature reservoirs, the residual current produces interfacial cooling.
Experimental feasibility is also discussed.
\end{abstract}

\maketitle

\section{Introduction}

Pressure-driven two-phase flow is commonly described by combining standard hydrodynamic laws with measured or modeled pressure drops, flow patterns, and heat-transfer coefficients \cite{Landau,Two-Phase-book}.
In particular, flow boiling in narrow channels has been studied intensively as a setting where liquid-gas interfaces control flow patterns, pressure losses, and heat-transfer characteristics \cite{Dhir,Thome,Karayiannis}.
Nevertheless, even in the simplest laminar geometry, it is not obvious how a pressure difference imposed by two reservoirs is distributed between the liquid bulk, the gas bulk, and the interfacial region.
The question is especially sharp in the weak-driving regime, where ordinary hydrodynamic transport laws should hold in each bulk, while the interfacial state itself is not fixed by those bulk laws.

Existing theories usually introduce additional physics at the interface.
For evaporating and condensing interfaces, nonequilibrium thermodynamic approaches have been used to discuss interfacial transport and entropy production \cite{BedeauxKjelstrupRubi}.
Evaporation-condensation analyses between parallel plates also combine liquid and vapor bulk transport with kinetic conditions at the interface \cite{Chen2023}.
Temperature jumps and thermal resistance at liquid-vapor interfaces have been reported experimentally and in molecular simulations \cite{FangWard99,Muscatello17}.
Sharp-interface methods close the problem by prescribing jump, kinetic, and surface-tension relations at the phase boundary \cite{Fechter}, while diffuse-interface and dynamic van der Waals descriptions resolve the interfacial layer through square-gradient free energies and the resulting capillary stresses \cite{vdWCapillarity,Korteweg,Onuki}.

The purpose of this paper is different.
We ask what follows before interfacial kinetic laws, jump conditions, or stress balances are imposed.
We therefore study Poiseuille flow of a one-component fluid between two adiabatic plates, with a liquid reservoir on the left and a gas reservoir on the right.
The two bulk regions are assumed to show Poiseuille flow and Fourier heat conduction in the linear response regime.
The particle current and the energy current are conserved across the system, but the pressure and temperature on the two sides of the interfacial region are kept as independent boundary values.
Thus the present theory does not select the microscopic interfacial state, the interface position, or the normal-stress balance at the interface.

The calculation below identifies extremely small dimensionless parameters $A^\subL$ and $A^\subG$ describing coexistence Poiseuille flow, whose smallness comes from squared microscopic-to-macroscopic length ratios.
In weak driving, when both liquid and gas bulk regions remain macroscopic, the imposed pressure difference is concentrated as an interfacial pressure gap.
Only small residual pressure drops remain to drive Poiseuille flow in the two bulks, resulting in momentum insulation, namely a strong reduction of the ordinary Poiseuille particle current.
The small residual particle current may also produce interfacial cooling.

No interfacial model is used to obtain these conclusions.
The present bulk theory does not determine the microscopic interfacial state, the interface position, or the stability of a flat transverse interface.
Rather, it identifies the pressure gap that interfacial physics must support, reduce, or accommodate.
Possible responses include capillarity and contact-line pinning, deformation or rearrangement of the two-phase flow pattern, diffuse-interface stresses, interfacial transport resistance, and the breakdown of continuum hydrodynamics at molecular scales \cite{Bonn,Ruspini,Anderson,SaurelPantano,Chen2022,Chen2024,KalempaGraur}.

\section{Formulation}

\subsection{Setup}
\begin{figure}[tb]
\begin{center}
\includegraphics[width=0.8\linewidth]{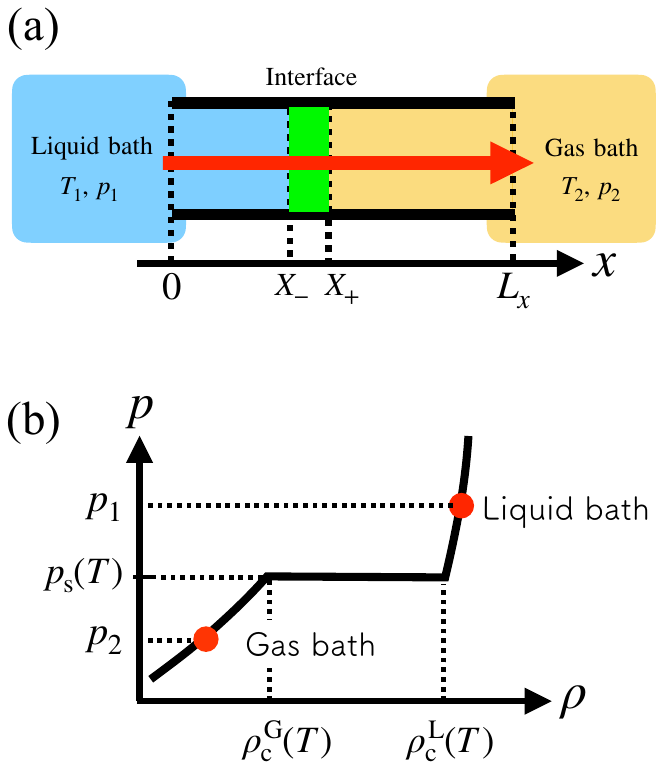}
\caption{(a) Schematic setup of liquid-gas Poiseuille flow between two particle reservoirs.
The bath variables are $(T_1,p_1)$ and $(T_2,p_2)$, and the interfacial region is represented by $[X_-,X_+]$.
(b) Reservoir phase conditions on a pressure--density isotherm for $T_1=T_2=T$.
The red points represent a liquid bath with $p_1>\ps(T)$ and a gas bath with $p_2<\ps(T)$.
The plateau endpoints $\rho_\subC^\subG(T)$ and $\rho_\subC^\subL(T)$ are the gas and liquid coexistence densities.}
\label{fig:setup}
\end{center}
\end{figure}

We consider laminar flow of a single-component fluid without gravity.
The fluid is isotropic with a molar mass $m$.
It flows between two adiabatic planes and is driven by two particle reservoirs as shown in Fig.~\ref{fig:setup}(a)
with a no-slip boundary condition at the surface of the planes.
The size of each plane is $L_x \times L_z$ and the distance between them is $L_y$.
We assume that $L_z$ is sufficiently large compared with $L_y$ that side-wall effects can be neglected.
The left reservoir in $x\le 0$ has pressure $p_1$ and temperature $T_1$,
and the right reservoir in $x\ge L_x$ has pressure $p_2$ and temperature $T_2$,
where we take $p_1 \ge p_2$. $T_1$ and $T_2$ are  below the critical temperature.

We focus on pressure-driven flow and define
\begin{align}
\ep\equiv\frac{p_1-p_2}{p},
\label{e:ep}
\end{align}
where $p$ is a representative pressure, such as $p_1$.
We assume $\ep\ll1$.
The bath temperatures need not be equal, but we require their relative difference $|T_1-T_2|/T$ to be at most of order $\ep$, where $T$ is a representative temperature, such as $T_1$.

The left and right reservoirs are maintained in the liquid and gas phases, respectively, each in a stable equilibrium state.
Their pressures satisfy $p_1\ge \ps(T_1)$ and $p_2\le \ps(T_2)$,
where $\ps(T)$ denotes the saturation pressure at temperature $T$.
Figure~\ref{fig:setup}(b) illustrates these phase conditions for equal-temperature baths.
The reservoir phases are maintained throughout the flow.

We restrict our analysis to the steady coexistence state reached after transient evolution, in which the interface is at rest and the liquid and gas bulk regions remain connected to their respective reservoirs.
Uniform translation of the interface would change the lengths of these regions and corresponds to a transient evolution in the finite channel.
A stationary interface is compatible with a nonzero particle current through phase change.

Let $(\rho, e,\pi^x, \pi^y, \pi^z)$ be the molar density, energy density, and three momentum densities, respectively,
and let the corresponding fluxes be
$({\bv j}_\rho,{\bv j}_{e},{\bv j}_{\pi}^x,{\bv j}_{\pi}^y,{\bv j}_{\pi}^z)$.
We have five continuity equations such as $\partial_t \rho+\nabla\cdot {\bv j}_{\rho}=0$.
Using the flow velocity field $\bv{v}(x,y,z)$ for $0<x<L_x$, $-L_y/2 <y<L_y/2$ and any $z$,
we have ${\bv \pi}=m\rho{\bv v}$ and $e=u+m\rho |\bv{v}|^2/2$ with internal energy density $u$.
The molar and energy fluxes are  written as $\bv{j}_\rho=\rho\bv{v}$, $\bv{j}_e=\hat h\bv{j}_{\rho}-\kappa\nabla T$, where $\hat h=(u+p)/\rho$ is the molar enthalpy and $\kappa$ is the thermal conductivity.
When the differences $p_1-p_2$ and $|T_1-T_2|$ are sufficiently small, the flow becomes  steady, laminar, and uniform in $z$.
We have the flow profiles as functions of $(x,y)$,  such as $\rho(x,y)$ for the molar density and
${\bv v}(x,y)$ for flow velocity.
Obviously, flow is parallel to $x$, i.e., $\bv{v}(x,y)=v(x,y)\bv{e}_x$ with unit vector $\bv{e}_x$  in the $x$-direction.
In this case,  the momentum fluxes are written as $\bv{j}_{\pi}^{x}=(p+m\rho{v}^2-\eta'\partial_x v)\bv{e}_x-\eta\partial_y v \bv{e}_y$, $\bv{j}_{\pi}^y=-\eta \partial_y v \bv{e}_x+p\bv{e}_y$, $\bv{j}_{\pi}^z=p\bv{e}_z$, where $\eta$ is shear viscosity and $\eta'=4\eta/3+\zeta$ with bulk viscosity $\zeta$.
Below, we concentrate on the currents  per unit area through each cross section at $x$, defined as
\begin{align}
{J}_{a}(x)\equiv \frac{1}{L_y}\inty \bv{e}_x\cdot {\bv j}_a(x,y),
\end{align}
where $a=(\rho, e,\pi^x, \pi^y, \pi^z)$.

\subsection{Single-phase bulk currents}

As preparation for the coexistence analysis, we first summarize the single-phase profiles and bulk current relations to be applied separately to the liquid and gas regions.
Solving the continuity equations $\nabla\cdot {\bv j}_{a}=0$ with the no-slip boundary condition, ${\bv v}(x, \pm L_y/2,z)=0$ for any $x$ or $z$
when $p_1-p_2$ and $T_2-T_1$ are sufficiently small,
we have
\begin{align}
&T(x,y)=T_1+\frac{T_2-T_1}{L_x}x+O(\ep^2)\equiv T(x), \label{e:T(x)} \\
&p(x,y)=p_1-\frac{p_1-p_2}{L_x}x+O(\ep^2)\equiv p(x), \label{e:p(x)}\\
&v(x,y)=\frac{p_1-p_2}{2\eta L_x}\left(\frac{L_y^2}{4}-y^2\right)+O(\ep^2)\equiv v(y), \label{e:v(y)}
\end{align}
where $\ep$ is  the pressure-driving parameter defined in \eqref{e:ep}.

We retain terms only to first order in $\ep$, with $v=O(\ep)$. Viscous heating and kinetic-energy transport are of second or higher order and are neglected.
The adiabatic boundary condition at the plates gives no transverse heat flux.
Note that $T$ and $p$ depend only on $x$ while $v$ depends only on $y$ in the linear response regime ignoring $O(\ep^2)$.
From the equation of state, the molar density is considered as a function of $x$ as
\begin{align}
\rho(x)=\rho(T(x),p(x))\label{e:rho(x)}
\end{align}
for any $y$.
We then have
\begin{align}
&J_\rho=-\frac{L_y^2}{12m\nu}\frac{p_2-p_1}{L_x}+O(\ep^2), \label{e:J_rho}\\
&J_e=\hat h J_\rho-\kappa \frac{T_2-T_1}{L_x}+O(\ep^2),\label{e:J_e}\\
&J_{\pi}^x(x)= p(x)+O(\ep^2), \label{e:J_pi(x)}\\
&J_{\pi}^y(x)= O(\ep^2), \label{e:J_piy(x)}
\end{align}
where ${\nu}\equiv{\eta}/({m\rho})$ is kinematic viscosity.
Since $J_{\pi}^y=O(\ep^2)$, we neglect it in the linear response regime,
and write $J_{\pi}^x(x)$ as $J_{\pi}(x)$ for simplicity.
In the linear response regime, $\hat h$, $\nu$, and $\kappa$ are considered constants, with $\hat h=\hat h(T,p)$, $\nu=\nu(T,p)$, and $\kappa=\kappa(T,p)$, where
$T$ and $p$ are representative  values of temperature and pressure, such as $T_1$ and $p_1$.
We note that the numerical factor $12$ in Eq.~\eqref{e:J_rho} is the Poiseuille coefficient for parallel plates.
The following arguments also apply to a circular pipe with radius $L_y$ by replacing this factor by $8$.
For a textbook treatment of viscous flow and heat conduction, see, for example, Landau and Lifshitz \cite{Landau}.

\subsection{Coexistence geometry and interfacial-side values}
We now proceed to liquid-gas coexistence in laminar flow.
We assume that the interfacial region occupies $X_-<x<X_+$.
Its thickness is denoted by $\Delta_X\equiv X_+-X_-$, and is assumed to be small compared with the channel length,
\begin{align}
\Delta_X\ll L_x.
\end{align}
A representative liquid-gas surface in this region is taken to be perpendicular to the $x$-axis.
The surface may be curved, but it remains confined inside the interfacial region.
The two adjacent liquid and gas bulk regions, $0<x<X_-$ and $X_+<x<L_x$, are assumed to remain macroscopic in the weak-driving states considered.
We require the interfacial region to be thin compared with each bulk, $\Delta_X\ll X_-$ and $\Delta_X\ll L_x-X_+$. In each bulk, the local equation of state and the transport laws of Sec.~II B are assumed to apply.
Below, superscripts $\subL$ and $\subG$ are attached to quantities related to liquid and gas, respectively.
For instance, the energy current is $J_e^\subL$ or $J_e^\subG$ for the liquid or gas region,
and the equation of state in liquid or gas is written as $\rho=\rho^\subL(T,p)$ or $\rho=\rho^\subG(T,p)$.
We denote the temperatures and pressures at the two boundaries of the interfacial region by
\begin{align}
&\theta_-\equiv \lim_{x\nearrow X_-}T(x),
\quad
\theta_+\equiv \lim_{x\searrow X_+}T(x),\\
&
p_-\equiv \lim_{x\nearrow X_-}p(x),
\quad
p_+\equiv \lim_{x\searrow X_+}p(x).
\end{align}
No condition is imposed here to relate these four interfacial-side values.
In particular, no normal-stress balance is imposed to identify $p_-$ and $p_+$.

\subsection{Bulk current relations in coexistence}

Particle conservation gives $J_\rho^\subL=J_\rho^\subG\equiv J_\rho$.
We can apply the bulk formulas for the currents, i.e.,
$J_\rho^\subLG$ is written  using \eqref{e:J_rho} as
\begin{align}
J_\rho
=
-\frac{L_y^2}{12m\nu^\subL}\frac{p_- -p_1}{X_-}
=
-\frac{L_y^2}{12m\nu^\subG}\frac{p_2-p_+}{L_x-X_+}.
\label{e:J_rho^LG}
\end{align}
Adding the two bulk pressure drops obtained from Eq.~\eqref{e:J_rho^LG}, we obtain
\begin{align}
p_1-p_-+p_+-p_2
&=
\frac{12m}{L_y^2}
\left[\nu^\subL X_-+\nu^\subG(L_x-X_+)\right]J_\rho.
\label{e:Jrho-pressure-sum}
\end{align}
Equivalently,
\begin{align}
J_\rho
=\frac{L_y^2}{12m\nu^{\rm eff}}\frac{p_1-p_-+p_+-p_2}{L_x-\Delta_X},
\label{e:J_rho-diff}
\end{align}
where the effective kinematic viscosity is
\begin{align}
\nu^{\rm eff}
=\frac{\nu^\subL X_-+\nu^\subG(L_x-X_+)}{L_x-\Delta_X}.
\label{e:nu_eff}
\end{align}
The numerator $p_1-p_-+p_+-p_2$ is the sum of the pressure drops in the two bulk regions.

Energy conservation gives $J_e^\subL=J_e^\subG\equiv J_e$.
Using the bulk energy-current formula \eqref{e:J_e} on the two sides, we have
\begin{align}
J_e=
-\kappa^\subL\frac{\theta_--T_1}{X_-}+\hat h^\subL J_\rho=-\kappa^\subG\frac{T_2-\theta_+}{L_x-X_+}+\hat h^\subG J_\rho.
\label{e:J_e^LG}
\end{align}
Adding the two bulk temperature changes obtained from Eq.~\eqref{e:J_e^LG} gives
\begin{align*}
T_2-T_1-\theta_++\theta_-
&=
-\left(
\frac{X_-}{\kappa^\subL}
+\frac{L_x-X_+}{\kappa^\subG}
\right)J_e
\\
&\quad+
\left(
\frac{\hat h^\subL X_-}{\kappa^\subL}
+\frac{\hat h^\subG(L_x-X_+)}{\kappa^\subG}
\right)J_\rho .
\end{align*}
Equivalently,
\begin{align}
&J_e=
-\kappa^{\rm eff} \frac{T_2-T_1-\theta_++\theta_-}{L_x-\Delta_X}+\hat h^{\rm eff} J_\rho,
\label{e:J_e-T}
\\
&\kappa^{\rm eff}=\frac{\kappa^\subL\kappa^\subG (L_x-\Delta_X)}{\kappa^\subL (L_x-X_+)+\kappa^\subG X_-},\label{e:kappa_eff}
\end{align}
and
\begin{align}
\hat h^{\rm eff}=\frac{\kappa^\subL (L_x-X_+)\hat h^\subG+\kappa^\subG X_-\hat h^\subL}{\kappa^\subL (L_x-X_+)+\kappa^\subG X_-}.
\label{e:p-gap}
\end{align}
The coefficient $\hat h^{\rm eff}$ is the effective enthalpy associated with the advective contribution to the energy current through the two bulk thermal resistances in series.
The temperature combination $T_2-T_1-\theta_++\theta_-$ is the sum of the temperature changes across the two bulk regions.

\begin{table*}[bt]
\begin{tabular}{|c||c|c|c|c|c|c|c|c|c|c|c|}\hline
Species & molar mass & $T$  & $\hat q$ & $\rho^\subL$ & $\rho^\subG$ & $\kappa^\subL$ & $\kappa^\subG$ & $\nu^\subL$ & $\nu^\subG$ & $A^\subL L_y^2$ [m$^2$] & $A^\subG L_y^2$ [m$^2$]\\ \hline
H$_2$O & $0.0180$   & $373.12$  & $4.07\times 10^4$ & $5.32\times 10^4$ & $33.2$ & $0.679$ & $2.51\times 10^{-2}$ & $2.94\times 10^{-7}$ & $2.05\times10^{-5}$& $2.94\times 10^{-16}$ & $7.57 \times 10^{-16}$ \\ \hline
Ar& $0.0400$  & 90.0 & $6.35\times 10^3$ & $3.45\times 10^4$ & $186$ & $0.122$ & $5.74\times 10^{-3}$ & $1.80\times 10^{-7}$ & $9.93\times 10^{-7}$ & $1.25\times 10^{-16}$ & $3.26\times 10^{-17}$ \\ \hline
\end{tabular}
\caption{Saturation properties in SI units for H$_2$O and Ar. Their critical temperatures are  $647$ K for H$_2$O, and  $151$ K for Ar. The data are taken from Ref.~\cite{NIST}.}
\label{tab:H2O-Ar}
\end{table*}

\section{Key dimensionless quantities}

\subsection{Residual pressure drop and interfacial temperature change}

We introduce a representative temperature $T$  satisfying
\begin{align}
\kappa^\subL\frac{T-T_1}{X_-}
=
\kappa^\subG\frac{T_2-T}{L_x-X_+}.
\label{e:J_e^LG*}
\end{align}
For equal-temperature baths, this definition gives $T=T_1=T_2$.
For unequal bath temperatures, $T$ is a weighted average of $T_1$ and $T_2$, with weights proportional to the bulk thermal conductances per unit area, $\kappa^\subL/X_-$ and $\kappa^\subG/(L_x-X_+)$.
Using this definition together with the equality of the two expressions in \eqref{e:J_e^LG}, we obtain
\begin{align}
T=\theta_X
+\hat q \frac{X_-(L_x-X_+)}{\kappa^\subL(L_x-X_+)+\kappa^\subG X_-}J_\rho.
\label{e:T-theta_X-J_rho}
\end{align}
Here, $\hat q$ is the latent heat per mole, $\hat q\equiv \hat h^\subG(T,\ps(T))-\hat h^\subL(T,\ps(T))$,
and
$\theta_X$ is a representative interface temperature,
\begin{align}
\theta_X=\frac{\kappa^\subL (L_x-X_+)\theta_-+\kappa^\subG X_-\theta_+}{\kappa^\subL (L_x-X_+)+\kappa^\subG X_-}.
\label{e:theta_X-X2}
\end{align}
For later use, we record the sign information contained in this relation.
The coefficient of $J_\rho$ in \eqref{e:T-theta_X-J_rho} is positive because $\hat q>0$.
Thus, $J_\rho>0$ implies $\theta_X<T$, whereas $J_\rho<0$ implies $\theta_X>T$.
Because $\theta_X$ is an internal weighted average of $\theta_-$ and $\theta_+$ in \eqref{e:theta_X-X2},
at least one of the two interfacial-side temperatures lies on the same side of $T$ as $\theta_X$.
At this stage, however, this is only a consequence of the bulk energy balance and the definition of $\theta_X$.
Its interpretation as interfacial cooling will be discussed  for equal and unequal bath temperatures in Sec.~\ref{sec:cooling}.

We now introduce two dimensionless quantities related to the degree of non-equilibrium.
One measures the interfacial temperature change, and the other measures the residual pressure drop left in the two bulk regions:
\begin{align}
&\ep_T\equiv \frac{T-\theta_X}{T},
\label{e:ep_T}
\\
&\ep_p\equiv\frac{p_1-p_-+p_+-p_2}{\ps(T) }.
\label{e:ep_p}
\end{align}
Thus, $\ep_p$ measures the pressure drop left in the two bulk regions,
not the externally imposed pressure difference.
We then find that $J_\rho$ has two expressions; from \eqref{e:J_rho-diff}
\begin{align}
&J_\rho=\ep_p \frac{\ps(T)L_y^2}{12m\nu^{\rm eff}(L_x-\Delta_X)},
\label{e:J_rho-diff2}
\end{align}
and from \eqref{e:T-theta_X-J_rho}
\begin{align}
&J_\rho=\ep_T \frac{T}{\hat q}
\frac{\kappa^\subL\kappa^\subG}{\kappa^{\rm eff}}
\frac{(L_x-\Delta_X)}{X_-(L_x-X_+)},
\label{e:J_q-diff2}
\end{align}
where we applied \eqref{e:kappa_eff}.
Comparing \eqref{e:J_rho-diff2} with \eqref{e:J_q-diff2}, we have
\begin{align}
\ep_T\frac{T}{\ps(T)}=\ep_p \frac{\hat q L_y^2}{12 m}\frac{\kappa^{\rm eff}}{\nu^{\rm eff}\kappa^\subL\kappa^\subG}\frac{X_-(L_x-X_+)}{(L_x-\Delta_X)^2}.
\label{e:epT-epP-0}
\end{align}
We refer to Eq.~\eqref{e:epT-epP-0} as the bulk current-balance constraint.
It states that the residual bulk pressure drop, measured by $\ep_p$, and the interfacial temperature change, measured by $\ep_T$, cannot be chosen independently in a steady two-phase flow.

\subsection{\texorpdfstring{Dimensionless quantities $A^\subL$ and $A^\subG$ as microscopic-to-macroscopic ratios}{Dimensionless quantities A L and A G as microscopic-to-macroscopic ratios}}

We now rewrite \eqref{e:epT-epP-0} so that the material combinations controlling it become explicit.
Writing $\ps'=d\ps/dT$, multiplying \eqref{e:epT-epP-0} by $\ps'$, and substituting \eqref{e:nu_eff} and \eqref{e:kappa_eff}, we obtain
\begin{align}
\ep_T\frac{d\ln \ps}{d\ln T}
=&
\ep_p
\frac{\ps'\hat q L_y^2}{12m}
\frac{X_-}
{\nu^\subL X_-+\nu^\subG(L_x-X_+)}
\nonumber\\
&\quad\times
\frac{L_x-X_+}
{\kappa^\subL(L_x-X_+)+\kappa^\subG X_-}.
\label{e:compat-before-A}
\end{align}
\eqref{e:compat-before-A} has been obtained only from the bulk particle- and energy-current balances.
We have not related the four interfacial-side values $p_\pm$ and $\theta_\pm$ by an additional condition such as temperature continuity $\theta_-=\theta_+$, local saturation $p_\pm=\ps(\theta_\pm)$, or a normal-stress balance.

Using $r_\nu\equiv\nu^\subG/\nu^\subL$,
\eqref{e:compat-before-A} is rewritten as
\begin{align}
&\ep_T~\frac{d\ln \ps}{d\ln T}
\nonumber\\
&=\ep_p~\frac{r_\nu X_-(L_x-X_+)}{[A^\subL r_\nu(L_x-X_+)+A^\subG X_-][r_\nu(L_x-X_+)+X_-]},
\label{e:epT-epP}
\end{align}
where the dimensionless quantities $A^\alpha$ are defined for $\alpha=\subL,\subG$ by
\begin{align}
A^\alpha&\equiv \frac{12m}{\ps'\hat q L_y^2}\kappa^\alpha \nu^\alpha,
\qquad \alpha=\subL,\subG.
\label{e:def-A^LG}
\end{align}
Applying the Clausius-Clapeyron relation to \eqref{e:def-A^LG}, we have
\begin{align}
&A^\alpha=\frac{12m T(\rho^\subL-\rho^\subG)}{(\hat q L_y)^2\rho^\subL\rho^\subG}\kappa^\alpha\nu^\alpha,
\qquad \alpha=\subL,\subG.
\label{e:def2-A^LG}
\end{align}
In \eqref{e:def-A^LG} and \eqref{e:def2-A^LG}, $\kappa^\alpha$, $\nu^\alpha$, and $\rho^\alpha$ denote values at the saturated reference state $(T,\ps(T))$ used in the linear-response approximation.

We next give a simple molecular order estimate for $A^\subG$.
When $\rho^\subL\gg \rho^\subG$, $A^\subG$ in \eqref{e:def2-A^LG} is approximately written in terms of saturated gas quantities as
\begin{align}
&A^\subG
\simeq
\frac{12m T}{(\hat q L_y)^2\rho^\subG}\kappa^\subG\nu^\subG.
\label{e:def2-A^LG-apprx}
\end{align}
Below, an asterisk subscript denotes a molecular quantity, such as $m_*$ for the mass per particle.
The thermal speed is estimated as $v_*\sim \sqrt{\kB T/m_*}$ with the Boltzmann constant $\kB$.

In this gas kinetic estimate, $l_*$ denotes the mean free path between molecular collisions.
The typical time scale of the particle motion is estimated as $\tau_*=l_*/v_*\sim l_*\sqrt{m_*/\kB T}$.

We retain the number density $\rho_*=\rho^\subG N_{\rm A}$ explicitly, where $N_{\rm A}$ is the Avogadro number. The thermal energy density is estimated as $u_*\sim \rho_*\kB T$.
The heat-diffusion estimate $u_*/\tau_*\sim\kappa T/l_*^2$
leads to $\kappa\sim u_* l_*^2/(\tau_* T)$.
Combining the above estimates, we obtain
\begin{align}
\text{}\quad
\kappa \sim \rho_*\kB l_*\sqrt{\frac{\kB T}{m_*}}.
\label{e:kappa-estimate}
\end{align}
Similarly, the momentum-diffusion estimate $\partial_t \pi\sim\nu \partial_x^2\pi$ gives
$\nu\sim l_*^2/\tau_*$, i.e.,
\begin{align}
\nu\sim l_*\sqrt{\frac{\kB T}{m_*}}.
\label{e:nu-estimate}
\end{align}

Using $m=N_{\rm A}m_*$, $\rho^\subG=\rho_*/N_{\rm A}$, and $\hat q=N_{\rm A}\hat q_*$, where $\hat q_*$ is the latent heat per molecule, we substitute \eqref{e:kappa-estimate} and \eqref{e:nu-estimate} into \eqref{e:def2-A^LG-apprx}. The number density cancels, giving
\begin{align}
A^\subG\sim
\left(\frac{\hat q_*}{\kB T}\right)^{-2}
\left(\frac{l_*}{L_y}\right)^2.
\label{e:A-estimate}
\end{align}
This estimate shows that $A^\subG$ is governed by the square of the microscopic-to-macroscopic length ratio $l_*/L_y$,
apart from the molecular latent-heat factor.
For macroscopic $L_y$, this ratio is extremely small.

For ordinary fluids,
$A^\subL/A^\subG=\kappa^\subL \nu^\subL/\kappa^\subG \nu^\subG$ is not expected to be singular.
Thus, $A^\subL$ is of the same small order as $A^\subG$.

Table \ref{tab:H2O-Ar} lists saturation properties and representative values of $A^\subL L_y^2$ and $A^\subG L_y^2$, calculated from the database \cite{NIST} for H$_2$O at the 1 atm saturation point, $T=373.12$ K, and argon at $T=90$ K, far from the critical temperatures.
For water at $L_y=1$ cm, \eqref{e:def2-A^LG} yields $A^\subL=2.936\times 10^{-12}$ and $A^\subG=7.574\times 10^{-12}$.
Even for micrometer plate separations, the values in Table \ref{tab:H2O-Ar} give $A^\alpha\lesssim 10^{-3}$.
Making $A^\alpha$ of order unity would require lengths outside the range where ordinary continuum Poiseuille flow is reliable.
The numerical values are consistent with the order estimate \eqref{e:A-estimate}.
In what follows, the small-$A$ regime means $A^\subL,A^\subG\ll 1$, and $O(A)$ denotes a quantity at most of the order of $A^\subL$ or $A^\subG$.
The approximation $\rho^\subL\gg\rho^\subG$ replaces the factor $1-\rho^\subG/\rho^\subL$ by unity in \eqref{e:def2-A^LG-apprx}, before the gas kinetic estimate.
The bulk current balances and the definitions \eqref{e:def-A^LG} and \eqref{e:def2-A^LG} retain the full density dependence.
For comparable liquid and gas densities within the coexistence setting of Sec.~II C, $A^\subL$ and $A^\subG$ are evaluated from these definitions using the full latent heat and transport coefficients.
Pressure localization and momentum insulation require both $A^\subL,A^\subG\ll1$ under the weak-driving and regular-response assumptions stated in Sec.~IV A.

\section{Results}

\subsection{Pressure-gap localization}
\begin{figure}[tb]
\begin{center}
\includegraphics[scale=0.43]{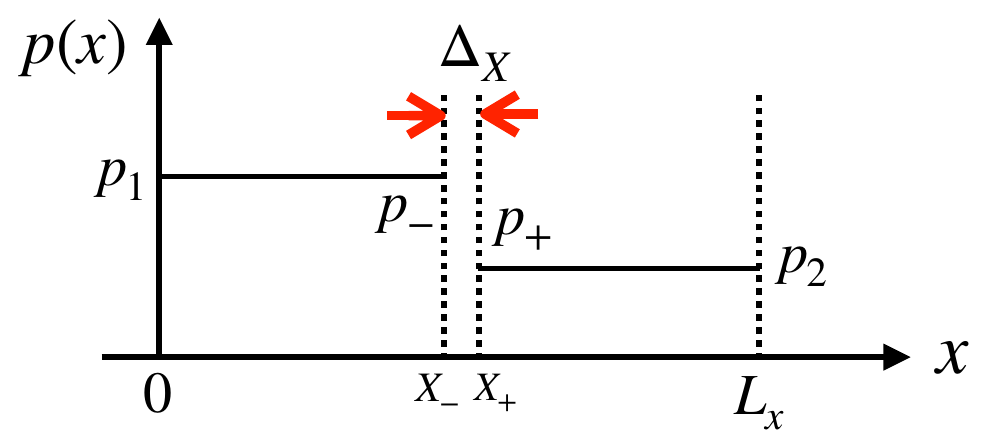}
\caption{Pressure profile in coexistence flow in the small-$A$ regime.
Only small residual pressure drops remain in the liquid and gas bulks, while most of the imposed pressure difference appears as a pressure gap across the interfacial region.}
\label{fig:profile}
\end{center}
\end{figure}

We now ask how the smallness of $A^\subL$ and $A^\subG$ affects the pressure distribution.
Equation \eqref{e:epT-epP} gives the scaling directly.
For ordinary fluids, $r_\nu$ is a material prefactor and does not compensate the smallness of $A^\alpha$.
Here, a regular interfacial temperature response means that $\ep_T$ remains  at most of the order of the imposed relative pressure difference $\ep$.
Thus $\ep_T=O(\ep)$ is the regular-response assumption, whereas the smallness of $A^\subL$ and $A^\subG$ comes from the material properties and the channel width. These are distinct small parameters; no ordering between $\ep$ and $A^\alpha$ is imposed in the first-order calculation.
When both $X_-$ and $L_x-X_+$ are macroscopic lengths, \eqref{e:epT-epP} gives
\begin{align}
\ep_p = O(\ep_T A),
\label{e:ep-eT}
\end{align}
which is expressed as
\begin{align}
\frac{p_1-p_-+p_+-p_2}{\ps(T)}=O(\ep_T A)
\label{e:result-p}
\end{align}
by substituting \eqref{e:ep_p}.
With $\ep_T=O(\ep)$, this gives $\ep_p=O(A\ep)$: the residual bulk pressure drop is suppressed by the small factor $A$ relative to the weak-driving scale.
Within the bulk Poiseuille and Fourier approximation, the bulk pressure drop remains proportional to $J_\rho$.
Combining this relation with the integrated bulk energy balance gives an $O(A\ep^2)$ correction to \eqref{e:ep-eT} for a nonzero latent heat, retaining the factor $A$ in this correction.
\eqref{e:result-p} shows that the pressure gap across the interfacial region is approximately $p_1-p_2$, while the pressure changes in the two bulk regions are negligible, as described in Fig.~\ref{fig:profile}.
Substituting \eqref{e:result-p} into \eqref{e:J_rho-diff},
we have
\begin{align}
\frac{12m\nu^{\rm eff}(L_x-\Delta_X)}{L_y^2\ps(T)}J_\rho=O(\ep_T A).
\label{e:insulation}
\end{align}
Thus the particle current driven by the residual bulk pressure drop is also as small as $O(\ep_T A)$.
Its magnitude relative to ordinary single-phase Poiseuille flow is discussed in Sec.~\ref{s:insulation}.

Under equal-temperature baths, $T_1=T_2=T$,  the two bulk pressures remain almost flat even though the reservoirs impose a pressure difference.
The state is nevertheless not an equilibrium state, because the residual particle current carries latent heat across the coexistence region.
The associated temperature signature is discussed in Sec.~\ref{sec:cooling}.

\subsection{Momentum insulation}
\label{s:insulation}

We now compare the particle current in phase coexistence with single-phase Poiseuille currents under the same imposed pressure difference $\Delta_p=p_1-p_2$.
For this comparison, take equal-temperature reservoirs, $T_1=T_2=T$.
The single-phase gas state corresponds to $p_2<p_1<\ps(T)$, and its particle current is denoted by $J_\rho^\subG$.
The single-phase liquid state corresponds to $p_1>p_2>\ps(T)$, and its particle current is denoted by $J_\rho^\subL$.
The coexistence state lies between them, $\ps(T)<p_1<\ps(T)+\Delta_p$, and its particle current is denoted by $J_\rho^{\rm LG}$.

Using the formulas \eqref{e:J_rho} and \eqref{e:J_rho-diff2}, we obtain the ratios of particle currents for fixed $\Delta_p$ as
\begin{align}
&\frac{J_{\rho}^{\rm LG}}{J_\rho^{\rm G}}
=\frac{r_\nu L_x}{X_-+r_\nu(L_x-X_+)}\frac{\ep_p ~\ps(T)}{\Delta_p}, \label{e:J_LG/G}\\
&\frac{J_{\rho}^{\rm L}}{J_\rho^{\rm G}}=r_\nu, \label{e:J_L/G}
\end{align}
where $r_\nu=\nu^\subG/\nu^\subL$.
The kinematic viscosity $\nu=\eta/(m\rho)$ depends on both the shear viscosity and the density.
The higher liquid density can outweigh the larger liquid shear viscosity, giving $\nu^\subL<\nu^\subG$.
The liquid can therefore carry a larger particle current at the same pressure difference even though its flow velocity is smaller.
The last factor in \eqref{e:J_LG/G} has a direct meaning.
From \eqref{e:ep_p},
\begin{align}
\frac{\ep_p\ps(T)}{\Delta_p}
=
\frac{p_1-p_-+p_+-p_2}{p_1-p_2},
\end{align}
 which is the fraction of the imposed pressure difference that remains as Poiseuille-driving pressure drops in the two bulk regions.
For the present equal-temperature weak-drive comparison, the regular-response assumption used in deriving \eqref{e:result-p} gives this fraction as $O(A)$ when both bulk lengths are macroscopic.
Therefore \eqref{e:J_LG/G} gives
\begin{align}
\frac{J_{\rho}^{\rm LG}}{J_\rho^{\rm G}}=O(A).
\label{e:J_LG-G-OA}
\end{align}
Because $A^\subL$ and $A^\subG$ contain microscopic-to-macroscopic length ratios, \eqref{e:J_LG-G-OA} means that the coexistence current is reduced by that small factor relative to the ordinary single-phase Poiseuille current.
Thus $J_\rho^{\rm LG}$ is controlled by the small residual pressure drop left in the two bulks, not by the full imposed pressure difference $\Delta_p$.

\begin{figure}[tb]
\begin{center}
\includegraphics[scale=0.45]{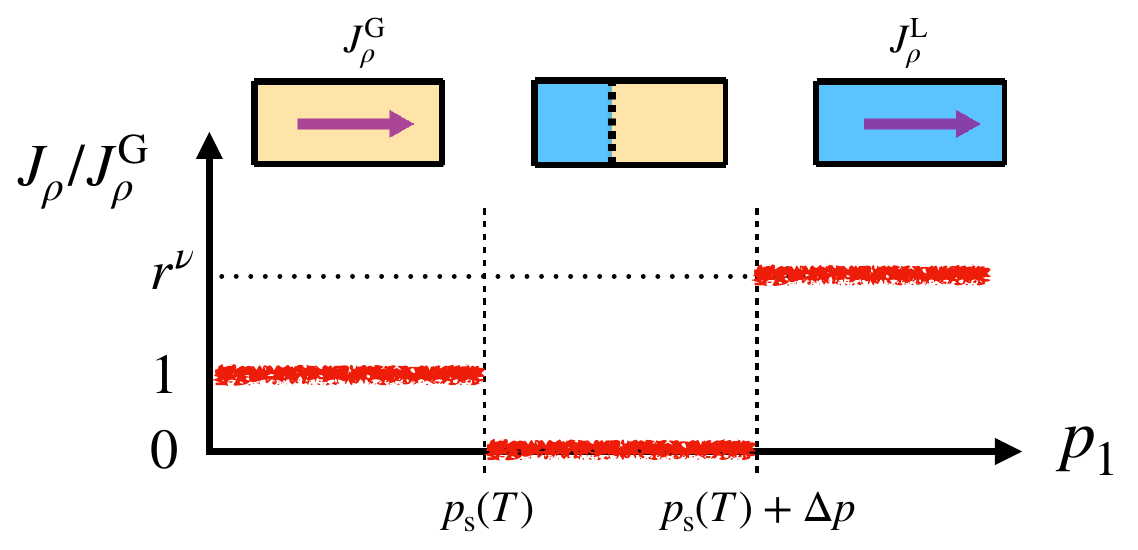}
\caption{Schematic particle current for fixed $\Delta_p=p_1-p_2$ as $p_1$ and $p_2$ are increased together.
The case $\nu^\subG>\nu^\subL$ is shown.
For states in the coexistence interval $\ps(T)<p_1<\ps(T)+\Delta_p$ with both bulk regions macroscopic, Eq.~\eqref{e:J_LG/G} gives a strongly reduced current compared with the single-phase gas Poiseuille current.}
\label{fig:flow-drop}
\end{center}
\end{figure}

Figure~\ref{fig:flow-drop} illustrates this comparison at fixed $\Delta_p$, while $p_1$ and $p_2$ are increased together.
In the interval $\ps(T)<p_1<\ps(T)+\Delta_p$, a liquid-gas interface is present.
For states away from the endpoints, with both bulk regions remaining macroscopic, \eqref{e:J_LG-G-OA} gives a strongly reduced particle current compared with the single-phase gas Poiseuille current.
This reduction follows from the bulk current-balance constraint, not from an additional phenomenological hydraulic resistance assigned to the interface.
In this restricted sense, the liquid-gas coexistence region acts as a momentum insulator.
We refer to the resulting reduction of the ordinary Poiseuille particle current as momentum insulation.

\subsection{Interfacial cooling from the residual particle current}
\label{sec:cooling}

We now examine the thermal consequence in the same equal-temperature setting, $T_1=T_2=T$.
Then the representative temperature introduced by \eqref{e:J_e^LG*} is the bath temperature $T$.
Equation \eqref{e:T-theta_X-J_rho} gives
\begin{align}
T-\theta_X
=
\hat q
\frac{X_-(L_x-X_+)}
{\kappa^\subL(L_x-X_+)+\kappa^\subG X_-}
J_\rho .
\label{e:cooling-Jrho}
\end{align}
Thus, for the pressure-driven state with $J_\rho>0$, the representative interfacial temperature satisfies
\begin{align}
\theta_X<T .
\end{align}
Because $\theta_X$ is the weighted average of $\theta_-$ and $\theta_+$ in \eqref{e:theta_X-X2},
at least one side of the interfacial region must be colder than the baths.

The bulk formulas also show when both sides are cooled.
From \eqref{e:J_e^LG} with $T_1=T_2=T$, one obtains
\begin{align}
&T-\theta_-
=
\frac{X_-}{\kappa^\subL}
\left(J_e-\hat h^\subL J_\rho\right),
\\
&T-\theta_+
=
\frac{L_x-X_+}{\kappa^\subG}
\left(\hat h^\subG J_\rho-J_e\right).
\label{e:theta-pm-Je}
\end{align}
Therefore both sides of the interfacial region are colder than the baths when
\begin{align}
\hat h^\subL J_\rho < J_e < \hat h^\subG J_\rho .
\label{e:Je-window}
\end{align}
This condition means that the energy current lies between the advective enthalpy currents of the liquid and gas phases.
When it holds, heat is conducted toward the interfacial region from both bulks according to Fourier's law.
If one additionally imposes $\theta_-=\theta_+$, this two-sided cooling follows immediately from $\theta_X<T$.

\begin{figure}[tb]
\begin{center}
\includegraphics[width=0.95\linewidth]{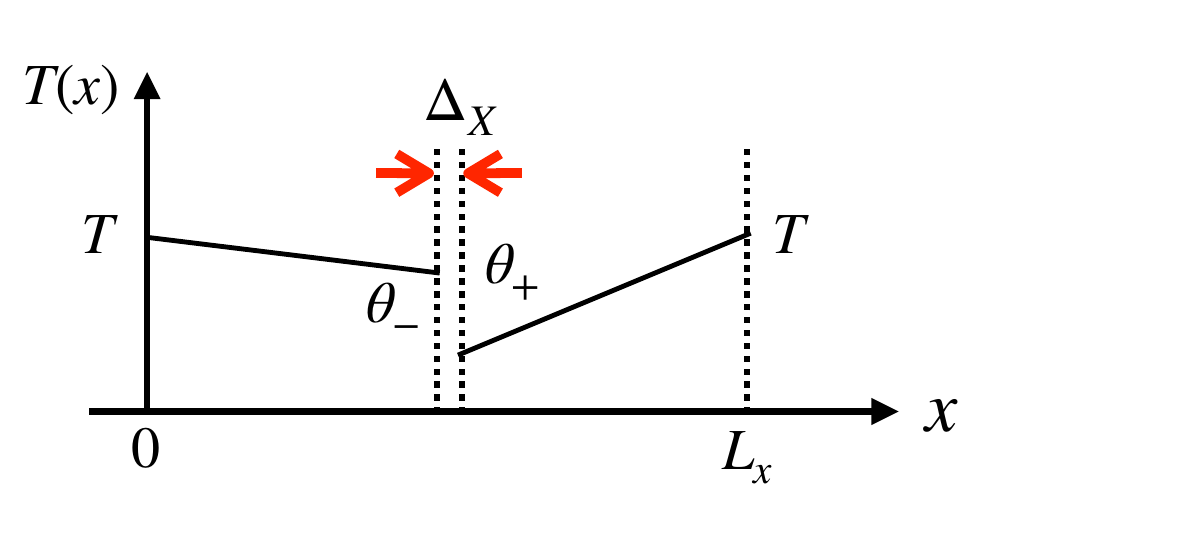}
\caption{Schematic temperature profile for interfacial cooling under equal-temperature baths.
The two-sided cooling drawn here corresponds to \eqref{e:Je-window}; for the pressure-driven state with $J_\rho>0$, \eqref{e:cooling-Jrho} gives $\theta_X<T$.}
\label{fig:diagram}
\end{center}
\end{figure}

This interfacial cooling is a thermal consequence of the residual particle current left by the bulk current-balance constraint.
Even though the two baths have the same temperature, a finite particle current carries latent heat through the interface.
In the two-sided case, this requires heat conduction from the bulk regions toward the cooled interfacial region, as indicated by the temperature profile in Fig.~\ref{fig:diagram}.
Temperature jumps and interfacial thermal resistance at liquid-vapor interfaces have been reported experimentally and in molecular simulations \cite{FangWard99,Muscatello17}.
The ordinary bulk equations alone do not determine whether the response appears as a temperature jump, a cooled interfacial layer, or deformation of the interface \cite{Chen2022,Chen2024,KalempaGraur}.

Reversing the current with liquid on the left and gas on the right would give $\theta_X>T$ from \eqref{e:T-theta_X-J_rho}.
At equal bath temperatures, however, the stable equilibrium reservoir conditions in Sec.~II A require $p_1\ge\ps(T)\ge p_2$, excluding a higher-pressure gas bath.

For unequal bath temperatures in the same liquid--gas arrangement, \eqref{e:T-theta_X-J_rho} gives $\theta_X<T$ for liquid-to-gas flow, whether $T_1<T_2$ or $T_1>T_2$.
Here $T$ is the weighted bath temperature defined by \eqref{e:J_e^LG*}; $\theta_X$ need not lie below the colder bath.
Beyond the pressure ordering $p_1\ge p_2$ adopted in Sec.~II A, the equilibrium reservoir conditions for $T_1<T_2$ also allow $p_2>p_1$.
If the current is reversed, \eqref{e:T-theta_X-J_rho} gives $T_1<T<\theta_X$, but $\theta_X$ need not exceed the hotter bath temperature $T_2$.
Possible changes in the phase arrangement when the bath temperatures are varied at fixed pressures are discussed in Appendix~\ref{app:phase-arrangement}.

\section{Experimental considerations}
\label{sec:cooling-mass-flux}

We consider water at $T_1=T_2=300\,\mathrm{K}$ in a horizontal adiabatic circular tube of diameter $d=1\,\mathrm{mm}$ and length $L_x=10\,\mathrm{cm}$, with $p_1-p_2=100\,\mathrm{Pa}$.
The millimeter diameter is chosen to limit heat transport by natural convection.
Liquid-water and pure-vapor reservoirs require $p_1\ge\ps(T)\ge p_2$, where $\ps(300\,\mathrm{K})\simeq3.54\,\mathrm{kPa}$.
We take a narrow interfacial region near the tube center, with $X_-\simeq L_x-X_+\simeq5\,\mathrm{cm}$.

\begin{figure}[tb]
\centering
\includegraphics[width=0.8\linewidth]{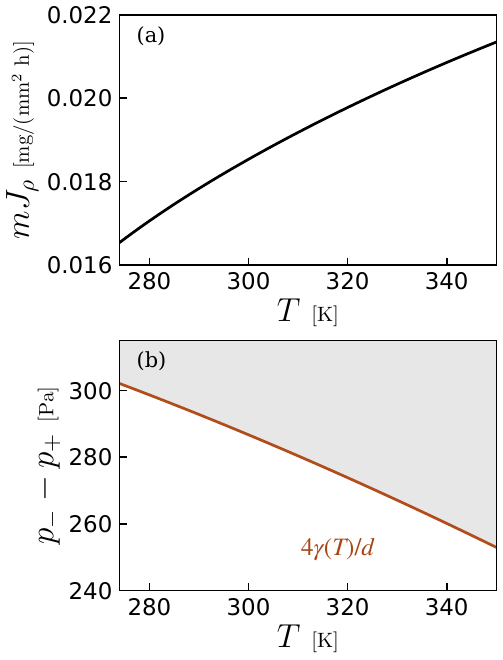}
\caption{Water in a circular tube with $d=1\,\mathrm{mm}$, two $5\,\mathrm{cm}$ bulks, and equal-temperature baths.
(a) Mass flux $mJ_\rho$ in $\mathrm{mg/(mm^2\,h)}$ for $T-\theta_X=1\,\mathrm{K}$, from \eqref{e:cooling-mass-flux}.
(b) Geometric Laplace-pressure bound $4\gamma(T)/d$ for a spherical-cap meniscus; gray shading marks values above the bound.
Saturated material properties are from Ref.~\cite{NIST}.}
\label{fig:cooling-mass-flux}
\end{figure}

The bulk energy balance \eqref{e:cooling-Jrho} also applies to a circular tube and gives
\begin{align}
mJ_\rho
=\frac{m(T-\theta_X)}{\hat q}
\left(\frac{\kappa^\subL}{X_-}
+\frac{\kappa^\subG}{L_x-X_+}\right).
\label{e:cooling-mass-flux}
\end{align}
The amount of mass current to observe the $1\, \mathrm{K}$ cooling from $T=300\,\mathrm{K}$ is estimated as $mJ_\rho\simeq0.0185\,\mathrm{mg/(mm^2\,h)}$ by substituting the saturation values in the NIST databook \cite{NIST}.
Figure~\ref{fig:cooling-mass-flux}(a) shows the temperature dependence for the same cooling and bulk lengths.

For a circular tube of a diameter $d$, $12/L_y^2$ in \eqref{e:Jrho-pressure-sum} is replaced by $32/d^2$, giving a total bulk pressure drop of $3.15\times10^{-3}\,\mathrm{Pa}$.
Nearly all the imposed pressure difference therefore lies across the interfacial region, $p_--p_+\simeq100\,\mathrm{Pa}$.

The maximum Laplace pressure magnitude is given by $4\gamma/d$, where $\gamma$ is the surface tension.
At $300\,\mathrm{K}$, this bound is $287\,\mathrm{Pa}$ [Fig.~\ref{fig:cooling-mass-flux}(b)], above the required $100\,\mathrm{Pa}$.

With mean velocity $\bar v^\alpha=J_\rho/\rho^\alpha$ and sound speed $c^\alpha$ in each phase $\alpha=\subL,\subG$, the Reynolds and Mach numbers are
\begin{align*}
\mathrm{Re}^\alpha
\equiv\frac{|\bar v^\alpha|d}{\nu^\alpha},
\qquad
\mathrm{Ma}^\alpha
\equiv\frac{|\bar v^\alpha|}{c^\alpha}.
\end{align*}
Using \eqref{e:ep}, \eqref{e:ep_T}, and \eqref{e:ep_p}, with $p=\ps(T)$, we obtain
\begin{gather*}
\ep\simeq2.83\times10^{-2},\;
\ep_T\simeq3.33\times10^{-3},\;
\ep_p\simeq8.91\times10^{-7},
\\
(\mathrm{Re}^\subL,\mathrm{Re}^\subG)
\simeq(6.03\times10^{-6},5.28\times10^{-4}),
\\
(\mathrm{Ma}^\subL,\mathrm{Ma}^\subG)
\simeq(3.44\times10^{-12},4.71\times10^{-7}).
\end{gather*}
Their smallness is consistent with the weak-driving bulk approximation.
Only a fraction $\ep_p/\ep\simeq3.15\times10^{-5}$ of the imposed pressure difference remains in the bulks.

\section{Summary and discussion}

We have studied pressure-driven Poiseuille flow of a one-component fluid in liquid-gas coexistence between adiabatic plates.
The analysis treats Poiseuille flow and Fourier heat conduction in the liquid and gas bulk regions together with particle and energy conservation.
No temperature-continuity condition, local-saturation condition, or normal-stress balance is imposed at the interface.
This bulk formulation yields two dimensionless quantities, $A^\subL$ and $A^\subG$, whose smallness originates from microscopic-to-macroscopic length ratios.

In the weak-driving regime with macroscopic liquid and gas bulk regions and a regular interfacial temperature response, the smallness of $A^\subL$ and $A^\subG$ localizes the pressure drop across the interfacial region.
Only an $O(A)$ fraction of the imposed pressure difference is left to drive Poiseuille flow in the two bulk regions.
Consequently, the coexistence particle current is $O(A)$ relative to ordinary single-phase Poiseuille flow.
In this sense, the liquid-gas coexistence region acts as a momentum insulator.
For equal-temperature baths, the remaining $O(A)$ particle current carries latent heat and produces interfacial cooling.

The main text assumes a liquid-gas coexistence state with a given interfacial position.
Appendix \ref{app:phase-arrangement} instead asks which candidate liquid-gas arrangements are suggested by the bath variables $(T_1,p_1)$ and $(T_2,p_2)$ before an interfacial position or a selection rule is specified.
Appendix \ref{app:forced-local-eq} asks what follows if, after the bulk calculation, one adds the stronger assumption of a zero-thickness saturated interface.
The latter comparison shows that this seemingly natural assumption becomes singular in the small-$A$ regime, pushing the interface toward the gas-bath side.
Neither appendix is used in deriving $A^\subL$ and $A^\subG$, nor does it select the steady state.

For the steady coexistence state considered here, the remaining questions concern how interfacial mechanics supports the localized pressure gap and how interfacial transport determines the current and temperature profile.

First, the interfacial position and shape are not fixed by the bulk balances.
The surface need not be flat: Sec.~II C allows a curved surface confined to a narrow region between the two macroscopic bulks.

Second, 
the localized pressure gap requires a mechanical balance across the interfacial region.
Capillarity and contact-line pinning provide one possible mechanism \cite{Bonn}.

The pressure scales for a narrow channel are discussed in Sec.~\ref{sec:cooling-mass-flux}.
If the pressure gap is supported solely by Laplace pressure, the bulk result $p_--p_+\simeq p_1-p_2$ constrains the interfacial curvature through the surface tension.
Determining the interfacial shape and position also requires the wall interactions and contact-line conditions.
The present bulk theory leaves these conditions unspecified, so identifying the mechanism that supports the pressure gap remains an open problem.

Third, the pressure variation may be carried by stresses inside a finite interfacial layer.
Diffuse-interface and square-gradient descriptions contain capillary stresses beyond the ordinary single-phase bulk stress and can represent pressure variation across a microscopic transition region \cite{DunnSerrin,Anderson,SaurelPantano}.

The present bulk calculation specifies the pressure gap that such an interfacial description must account for.

Fourth, the thermal response may be governed by interfacial transport processes not contained in the bulk Poiseuille-flow and Fourier-heat-conduction description, as in nonequilibrium thermodynamic descriptions of evaporating and condensing interfaces \cite{BedeauxKjelstrupRubi}.
In the equal-temperature case, Eq.~\eqref{e:cooling-Jrho} gives interfacial cooling produced by the residual particle current and latent-heat transport, not by a bath-temperature difference.
Interfacial thermal resistance, evaporation-condensation kinetics, and accommodation effects  may determine the temperature jump and the temperature profile within the interfacial region \cite{FangWard99,Muscatello17,Chen2022,Chen2024,KalempaGraur}.

The bulk energy balance relates the cooling to the current without determining these interfacial properties.
Measurements of stress and temperature differences, together with particle and energy currents, may guide a thermodynamic description of the region of width $\Delta_X$.
Existing interfacial thermodynamics \cite{BedeauxKjelstrup2005} provides a point of comparison.

Determining the steady current, interfacial position, and temperature profile for given bath conditions requires coupling the bulk balances to an interfacial description.

\appendix
\section{Candidate arrangements of liquid and gas}
\label{app:phase-arrangement}

\begin{figure}[tb]
\begin{center}
\includegraphics[width=0.95\linewidth]{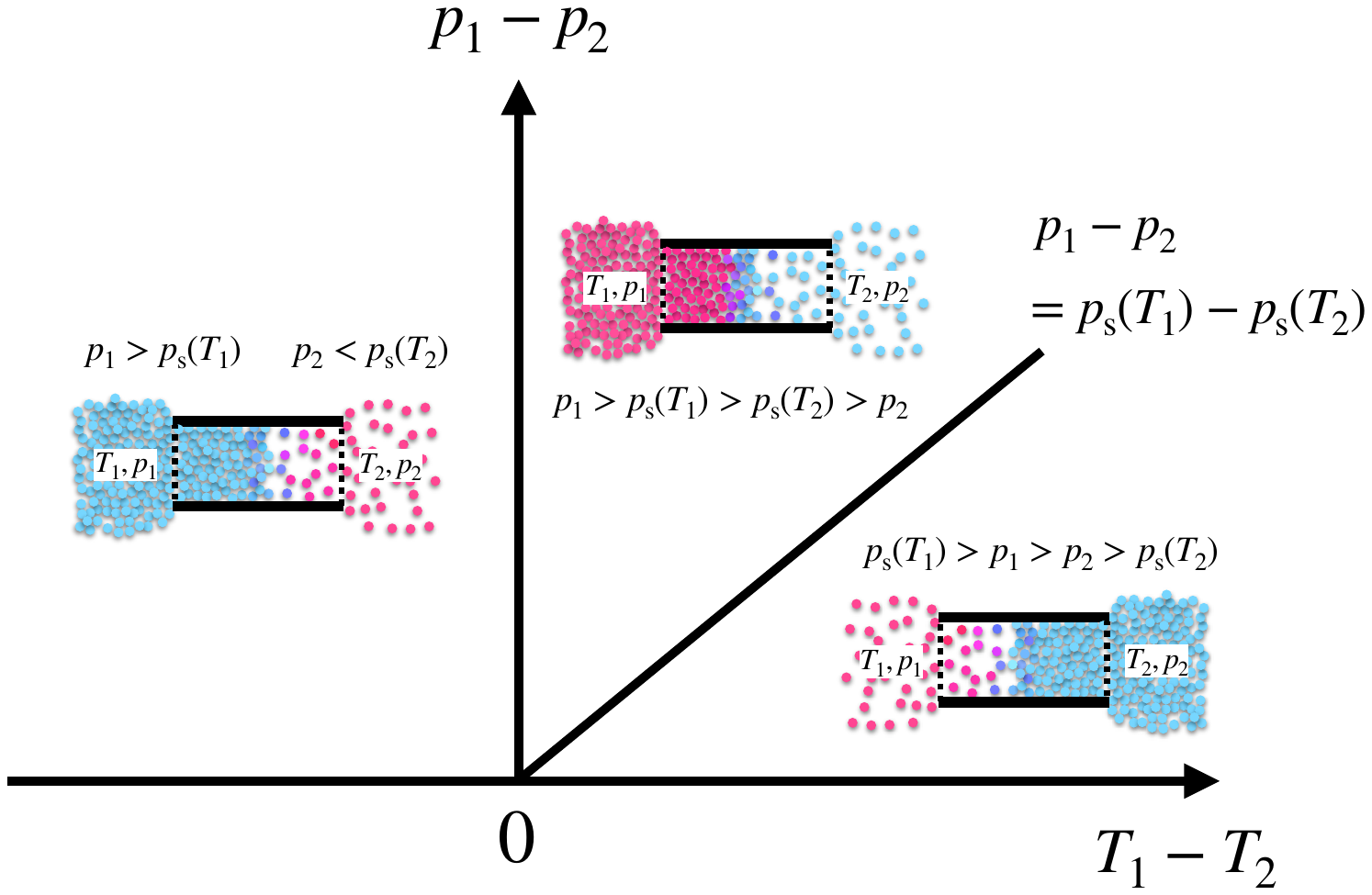}
\caption{Schematic phase-arrangement guide in the plane of bath differences $(T_1-T_2,p_1-p_2)$.
The diagonal line represents $p_1-p_2=\ps(T_1)-\ps(T_2)$.
The inequalities near the sketches show representative orderings of the bath pressures relative to the saturation pressures.
The sketches show candidate liquid-gas arrangements suggested by these comparisons.}
\label{fig:phase-arrangement}
\end{center}
\end{figure}

For bath 1, the comparison between $p_1$ and $\ps(T_1)$ indicates which phase is locally favored:
$p_1>\ps(T_1)$ favors the liquid phase, whereas $p_1<\ps(T_1)$ favors the gas phase.
The same comparison at bath 2 uses $p_2$ and $T_2$.
A natural boundary between different candidate arrangements is therefore obtained when the two deviations from saturation are equal,
$p_1-\ps(T_1)=p_2-\ps(T_2)$, or equivalently
\begin{align}
p_1-p_2=\ps(T_1)-\ps(T_2).
\label{e:phase-arrangement-line}
\end{align}
Together with these two local comparisons, \eqref{e:phase-arrangement-line} motivates the phase-arrangement guide in Fig.~\ref{fig:phase-arrangement}.

The inequalities shown in Fig.~\ref{fig:phase-arrangement} should be read only as labels attached to the two baths.
They clarify, for example, that the higher-temperature bath can still favor the liquid phase when its pressure exceeds the saturation pressure at its own temperature.
The balance line in \eqref{e:phase-arrangement-line} also appears in heat-conducting liquid-gas coexistence under gravity, where it gives the configurational inversion condition in a variational formulation of global thermodynamics \cite{NSgeffPRL26,NSgeff26}.
This connection suggests that \eqref{e:phase-arrangement-line} may be a useful starting point for a future variational principle for pressure-driven coexistence.
In the present paper, however, Fig.~\ref{fig:phase-arrangement} is used only as a guide to candidate arrangements.

\section{Saturated sharp-interface comparison}
\label{app:forced-local-eq}

The main text keeps $p_-$, $p_+$, $\theta_-$, and $\theta_+$ as independent interfacial-side values.
The bulk equations then give the bulk current-balance constraint \eqref{e:epT-epP-0}, but not an algebraic equation for a single interface pressure.
For comparison, we now collapse the interfacial region to a zero-thickness saturated surface.
That is, we set $\Delta_X=0$, $X_-=X_+=X$,
$\theta_-=\theta_+\equiv\theta_X$,
$p_-=p_+\equiv p_X$, and $p_X=\ps(\theta_X)$.
This assumption is not used in deriving \eqref{e:result-p} and \eqref{e:insulation}.
It also removes the possible pressure gap across the interfacial region and does not specify the normal-stress balance.
For simplicity we set $T_1=T_2=T$.

With this added assumption, the same bulk current relations \eqref{e:J_rho^LG} and \eqref{e:T-theta_X-J_rho} can be combined into one equation for $p_X$.
Linearizing $\ps(\theta_X)$ around $T$ and eliminating $X$, $J_\rho$, and $\theta_X$, we obtain
\begin{align}
p_X=\ps(T)
-\frac{(p_1-p_X)(p_X-p_2)}
{A^\subL(p_X-p_2)+A^\subG(p_1-p_X)} .
\label{e:eqn-p_X-A}
\end{align}
Equation \eqref{e:eqn-p_X-A} is a comparison result obtained only after adding the saturated sharp-interface relation $p_X=\ps(\theta_X)$ to the bulk current relations.

We now take the physical small-$A$ limit directly, without using the full algebraic solution.
We consider the branch with $p_1>\ps(T)>p_2$ and $p_1>p_X>p_2$.
Equation \eqref{e:eqn-p_X-A} gives $p_X<\ps(T)$.
If both $p_1-p_X$ and $p_X-p_2$ remained finite as $A^\subL,A^\subG\to0$, the second term in \eqref{e:eqn-p_X-A} would diverge and could not balance the finite pressures.
The branch near $p_X=p_1$ cannot satisfy $p_X<\ps(T)$.
Therefore the small-$A$ balance forces $p_X-p_2$ to be first order in $A^\subL$ and $A^\subG$.
Writing $p_X=p_2+b$ with $b=O(A)$, the leading balance of \eqref{e:eqn-p_X-A} gives $b=[\ps(T)-p_2]A^\subG+O(A^2)$.
Using this result in the sharp-interface form of \eqref{e:J_rho^LG} gives
\begin{align}
p_X&=p_2+[\ps(T)-p_2] A^\subG+O(A^2),
\nonumber\\
\frac{X}{L_x}
&=1-\frac{\ps(T)-p_2}{p_1-p_2}
\frac{\nu^\subL}{\nu^\subG}A^\subG+O(A^2),
\nonumber\\
\theta_X&=T_\subC(p_2+[\ps(T)-p_2] A^\subG+O(A^2)).
\label{e:sharp-smallA}
\end{align}
Here $O(A^2)$ denotes terms at least quadratic in $A^\subL$ and $A^\subG$, and $T_\subC(p)$ is the coexistence temperature, i.e., the inverse of $p=\ps(T)$.
The leading terms place $p_X$ close to $p_2$ and $X$ close to $L_x$.
They also give $\theta_X\simeq T_\subC(p_2)$.
On the branch considered here, $p_2<\ps(T)$, and therefore $T_\subC(p_2)<T$.

Thus the saturated sharp-interface assumption does not give a coexistence state with both liquid and gas bulk lengths of order $L_x$.
It instead drives the interface toward the gas-bath side and leaves an interfacial temperature decrease not reduced by the small parameters $A^\alpha$.
At exact equilibrium, $p_1=p_2=\ps(T)$, the interface position is not fixed by the bulk equations; the location found above is a singular consequence of imposing the saturated zero-thickness interface under a nonzero drive.
For this reason, the saturated sharp-interface comparison is not used as the steady-state selection rule in the main text.
The main result of the paper is instead the bulk current-balance constraint leading to \eqref{e:result-p} and \eqref{e:insulation}.

\begin{acknowledgments}
This work was supported by JSPS KAKENHI Grant Numbers JP23K22415, JP25K00923, JP25K22002, JP25H01975, and JP26H00383.
\end{acknowledgments}

\section*{Data Availability}
The thermophysical property data that support the numerical estimates in this article are openly available in Ref.~\cite{NIST}.
The remaining results are derived analytically in the paper.


\begin{thebibliography}{99}
\enlargethispage{\baselineskip}

\bibitem{Landau}
L. D. Landau and E. M. Lifshitz,
{\it Fluid Mechanics}
(Pergamon Press, Oxford, 1959).

\bibitem{Two-Phase-book}
S. M. Ghiaasiaan,
{\it Two-Phase Flow, Boiling, and Condensation in Conventional and Miniature Systems}, 2nd ed.
(Cambridge University Press, Cambridge, 2017).

\bibitem{Dhir}
V. K. Dhir,
Boiling heat transfer,
Annu. Rev. Fluid Mech. {\bf 30}, 365--401 (1998).

\bibitem{Thome}
J. R. Thome,
Boiling in microchannels: a review of experiment and theory,
Int. J. Heat Fluid Flow {\bf 25}, 128--139 (2004).

\bibitem{Karayiannis}
T. G. Karayiannis and M. M. Mahmoud,
Flow boiling in microchannels: Fundamentals and applications,
Appl. Therm. Eng. {\bf 115}, 1372--1397 (2017).

\bibitem{BedeauxKjelstrupRubi}
D. Bedeaux, S. Kjelstrup, and J. M. Rubi,
Nonequilibrium translational effects in evaporation and condensation,
J. Chem. Phys. {\bf 119}, 9163--9170 (2003).

\bibitem{Chen2023}
G. Chen,
On paradoxical phenomena during evaporation and condensation between two parallel plates,
J. Chem. Phys. {\bf 159}, 151101 (2023).

\bibitem{FangWard99}
G. Fang and C. A. Ward,
Temperature measured close to the interface of an evaporating liquid,
Phys. Rev. E {\bf 59}, 417--428 (1999).

\bibitem{Muscatello17}
J. Muscatello, E. Chac\'on, P. Tarazona, and F. Bresme,
Deconstructing temperature gradients across fluid interfaces:
The structural origin of the thermal resistance of liquid-vapor interfaces,
Phys. Rev. Lett. {\bf 119}, 045901 (2017).

\bibitem{Fechter}
S. Fechter, C.-D. Munz, C. Rohde, and C. Zeiler,
A sharp interface method for compressible liquid-vapor flow with phase transition and surface tension,
J. Comput. Phys. {\bf 336}, 347--374 (2017).

\bibitem{vdWCapillarity}
J. D. van der Waals,
The thermodynamic theory of capillarity under the hypothesis of a continuous variation of density,
translated by J. S. Rowlinson,
J. Stat. Phys. {\bf 20}, 197--244 (1979),
doi:10.1007/BF01011513.

\bibitem{Korteweg}
D. J. Korteweg,
Sur la forme que prennent les {\'e}quations du mouvement des fluides si l'on tient compte des forces capillaires caus{\'e}es par des variations de densit{\'e},
Arch. N{\'e}erl. Sci. Exactes Nat. {\bf 6}, 1--24 (1901).

\bibitem{Onuki}
A. Onuki,
Dynamic van der Waals theory of two-phase fluids in heat flow,
Phys. Rev. Lett. {\bf 94}, 054501 (2005).

\bibitem{Bonn}
D. Bonn, J. Eggers, J. Indekeu, J. Meunier, and E. Rolley,
Wetting and spreading,
Rev. Mod. Phys. {\bf 81}, 739--805 (2009).

\bibitem{Ruspini}
L. C. Ruspini, C. P. Marcel, and A. Clausse,
Two-phase flow instabilities: a review,
Int. J. Heat Mass Transfer {\bf 71}, 521--548 (2014).

\bibitem{Anderson}
D. M. Anderson, G. B. McFadden, and A. A. Wheeler,
Diffuse-interface methods in fluid mechanics,
Annu. Rev. Fluid Mech. {\bf 30}, 139--165 (1998).

\bibitem{SaurelPantano}
R. Saurel and C. Pantano,
Diffuse-interface capturing methods for compressible two-phase flows,
Annu. Rev. Fluid Mech. {\bf 50}, 105--130 (2018).

\bibitem{Chen2022}
G. Chen,
On the molecular picture and interfacial temperature discontinuity during evaporation and condensation,
Int. J. Heat Mass Transfer {\bf 191}, 122845 (2022).

\bibitem{Chen2024}
G. Chen,
Interfacial cooling and heating, temperature discontinuity and inversion in evaporation and condensation,
Int. J. Heat Mass Transfer {\bf 218}, 124762 (2024).

\bibitem{KalempaGraur}
D. Kalempa and I. Graur,
Temperature and pressure jump coefficients at a liquid-vapor interface,
Phys. Fluids {\bf 36}, 083622 (2024).

\bibitem{NIST}
E. W. Lemmon, I. H. Bell, M. L. Huber, and M. O. McLinden,
Thermophysical Properties of Fluid Systems,
in \textit{NIST Chemistry WebBook, NIST Standard Reference Database Number 69},
eds. P. J. Linstrom and W. G. Mallard,
National Institute of Standards and Technology,
Gaithersburg MD, 20899,
\url{https://doi.org/10.18434/T4D303},
retrieved May 24, 2026.

\bibitem{DunnSerrin}
J. E. Dunn and J. Serrin,
On the thermomechanics of interstitial working,
Arch. Rational Mech. Anal. {\bf 88}, 95--133 (1985).

\bibitem{NSgeffPRL26}
N. Nakagawa and S.-i. Sasa,
Thermodynamic variational principle unifying gravity and heat flow,
Phys. Rev. Lett. {\bf 136}, 057101 (2026),
doi:10.1103/bbqy-hptc.

\bibitem{NSgeff26}
N. Nakagawa and S.-i. Sasa,
Global thermodynamics for heat-conducting fluids under weak gravity,
arXiv:2605.31233 (2026).

\bibitem{BedeauxKjelstrup2005}
D. Bedeaux and S. Kjelstrup,
Heat, mass and charge transport, and chemical reactions at surfaces,
Int. J. Thermodynamics {\bf 8}, 25--41 (2005).

\end{thebibliography}
\end{document}